\documentclass[fleqn,twoside]{article}
\usepackage{espcrc2}

\input BoxedEPS
\SetRokickiEPSFSpecial  %% for dvips by Tom Rokicki, for VMS
\HideDisplacementBoxes

\title{Opportunities, Challenges, and Fantasies  in Lattice QCD}

\author{Frank Wilczek\address[CTP]{Center for Theoretical Physics\\
Massachusetts Institute of Technology\\
Cambridge, MA 02139-4307}%
        \thanks{The research is
supported in part by funds provided by the U.S. Department of Energy (D.O.E.)
under cooperative research agreements Nos.\ DE-FC02-94-ER40818 and
DE-FG02-91-ER40676.
\quad  MIT-CTP-3337.
}}

\begin{document}

\begin{abstract}\noindent
Some important problems in quantitative QCD  will certainly yield to hard
work and adequate investment of resources, others  appear difficult but may be
accessible, and still others  will require essentially new ideas.   Here I
identify
several examples in each class.
%\vspace{1pc}
\end{abstract}

\maketitle

There has been a notable  renaissance of interest and progress in QCD over the
last few years.  This has come  about for many reasons, including
\begin{itemize}
\item a vigorous experimental program in heavy ion physics explicitly
devoted to
exhibiting the fundamental dynamics of quark and gluon degrees of freedom and
to recreating conditions last seen in the Universe at $\sim 10^{-2}$~s
following
the Big Bang,
\item a vigorous experimental program in heavy quark physics, including
measurements of fundamental CP violation and weak mixing parameters, that
requires accurate theoretical calculation of strong matrix elements to
reach its full
potential,
\item creative developments in finite temperature theory, effective field
theory,
and numerical algorithms that dovetail beautifully with  these experimental
programs,
\item realization that high density is a regime of QCD in which weak
coupling but
nonperturbative methods can be used to give tractable yet rigorous models of
confinement and chiral symmetry breaking, with possible application to neutron
star interiors (or quark stars!? or strangelets!?)
\item development of lattice regularizations that incorporate chiral symmetry,
\item incremental developments in algorithms and hardware that have now
matured to the extent that lattice gauge theory has become a powerful tool
capable of providing reliable quantitative information genuinely useful for
guiding
and interpreting  experimental work, and
\item increasing anticipation of new frontier hadronic accelerators, the
upgraded
Tevatron, and especially the Large Hadron Collider (LHC), where QCD will
dominate
both input and output.  It will provide a challenging ``background''
against which
to analyze and interpret any essentially new phenomena.
\end{itemize}

Lattice gauge theory has been an essential ingredient in this ferment. This
subject  now appears, I believe, more promising and more important than
ever before.  It is blessed with many attractive research programs on
various time scales.  We can classify these roughly into:
\begin{itemize}
\item opportunities -- significant problems that will almost certainly yield to
hard work and adequate investment of resources in a reasonable, predefined
period of time,
\item challenges --  more ambitious problems, which may or may not yield to
incremental development of known techniques, and
\item fantasies -- grand problems that we can articulate, but presently lack
usable tools to address.
\end{itemize} I've had fun using this framework to think strategically
about the
future, in general.   I'll use it here to touch on three major branches of
lattice gauge
theory -- few-body, many-body, and conceptual/algorithmic aspects -- by
providing a couple of examples in each category.

\section{Fundamental Particle Physics}

\subsection{Opportunities}

\medskip

\subsubsection{The Best Determination of \protect\boldmath$\alpha_s$}

A major triumph of lattice QCD is already enshrined in the Particle Data
Book, as
shown in Figure 1.  Nonperturbative calculations of heavy quark spectroscopy
provide a precision determination of
$\alpha_s$ competitive with the most accurate determinations from perturbative
QCD.   There is probably more room for improvement on the nonperturbative
side, because on that side the relation of the calculations to observables
are in principle more tightly
controlled (no structure functions, jet parameterizations, ... ), the
experimental measurements are very
precise, and one is accessing $\alpha_s$ (or
$\alpha_s^p, p>0$) effects directly, rather than digging them out as
corrections.
It is important to pursue this direction further, for several reasons.

\begin{figure}[ht]
\centerline{\BoxedEPSF{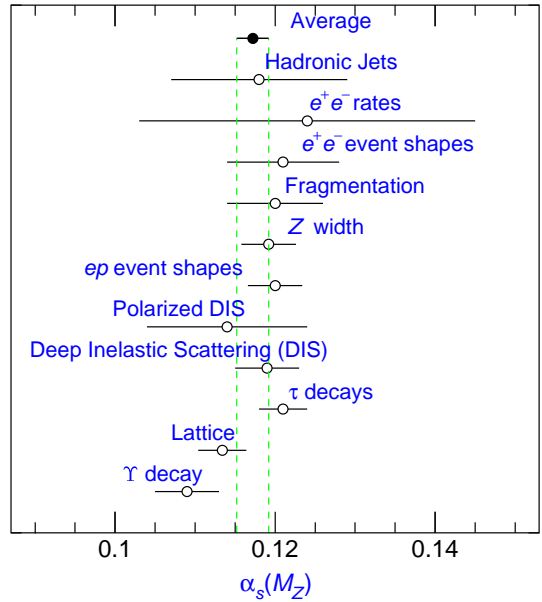 scaled 650}}
\vspace*{-\bigskipamount}
\caption{Leading determinations of the strong coupling $\alpha_s(M_Z)$, taken
from the Particle Data Book.}
\label{F1}
\vspace*{-\bigskipamount}
\end{figure}

Most obviously, a better determination of $\alpha_s$, by sharpening input to the
perturbative framework, supports more precise predictions for high-energy
experiments.   Also, if we are to have confidence in using lattice QCD  as a
mathematical tool for determining weak matrix elements, we need to make
sure it
gives consistent, accurate results for a variety of quantities in
spectroscopy and
electromagnetic transitions, where the underlying physics is securely
known.   These quantities, of course,
are just those that go into the lattice QCD determination of $\alpha_s$.

Precision determination of $\alpha_s$ will also add to the power of the
unification of
couplings calculation, which at present provides our best quantitative
handle on
fundamental physics beyond the standard model.   The central
value is 
on the low side for supersymmetric grand unified theories, but perhaps within
the margin of plausible threshold corrections.    If and when we start to
pin down
the low-energy supersymmetric spectrum, we will be able to leverage this
precision to gain insight into how unification symmetry and supersymmetry are
broken.

Finally, I think it's very legitimate to regard precise determination of
$\alpha_s$
as an end in itself.   Together with the electron mass, the ordinary fine
structure
constant $\alpha$, and the up and down quark masses,  $\alpha_s$ is one of
a handful of
parameters that determines the structure of ordinary matter.    It is a unique
glory of physics that we demand {\it precise, quantitative\/} agreement between
our calculations and reality, whenever comparison is possible.  Enough said.

\subsubsection{Exploring Canonical Structures}

%few parameters $\to$ canonical structures

The fact that QCD is a tight, conceptually defined theory implies that the
structures it produces are canonical.

Let me explain what I mean by this, by way of an analogy with black hole
physics.
A feature of black holes that makes them especially
fascinating and attractive is that their theory is so clean.  The key to
this cleanliness is that, as Wheeler put it,
``black holes have no hair''.    Specifically, classical
general relativity produces unique
predictions for the properties of a black hole, given only its mass and
angular momentum.    This is
quite different from  stars or ordinary matter, whose structure depends on many
parameters including elemental composition, temperature, and processing
history.   The canonical, nearly parameter-free structure of black holes
means that it is worthwhile to
study solutions of the equations that govern them minutely, because these
solutions are sharply delimited.

In QCD the structure of all the hadrons is canonical, and the structure of
protons
and neutrons is especially so since they are stable particles, i.e.,
discrete eigenstates of the Hamiltonian.
Nucleons are much balder than black holes,  for they have no hair, not only
classically but even in
quantum theory.  Furthermore, there is no freedom to specify their mass and
angular momentum arbitrarily.   Also protons and
neutrons can be well modeled in QCD
Lite$^{\rm TM}$ (the idealization of QCD using only massless $u$
and $d$
quarks).   In that approximation, they are entirely conceptually determined
structures, allowing no continuous parameters whatsoever!

And they can be accessed
experimentally.
Indeed, the answer to many ``practical'' questions relies on understanding
their
structure.  For example, structure functions are vital to the
interpretation of accelerator experiments, since we use nucleons as
projectiles.
Notably, the primary mechanism for Higgs particle production at the LHC will be
gluon fusion,
$gg \rightarrow h$ through a top-quark loop.   To predict the rates, and to
check
for possible deviations from standard model expectations for the $h$
coupling, we
obviously need to know the gluon distribution in the proton accurately.

Another sort of canonical structure in QCD is the flux tube between heavy quark
sources.  There are very interesting questions concerning its spatial
structure and
modes of excitation, and specifically how accurately it can be
modeled as a
bag or an elementary string.

\subsection{Challenges}\medskip

\subsubsection{Precision Values of the\\ Quark Masses}
\medskip
{\bfseries\boldmath$m_u$ and \boldmath$m_d$\qquad}In contrast to the beautiful
and inspiring situation regarding
$\alpha_s$, our present determinations of $m_u$ and
$m_d$ are an embarrassment.   They are uncertain at the factor-of-two level, at
least.  Despite their crucial importance  for the structure of the
world\footnote{See, in this connection, the following discussion of
quantitative
anthropics in Section 1.3.2.}, they are by far the worst measured fundamental parameters of
physics.    It is a major challenge for lattice gauge theory to improve this
situation.

Again, a comparison may be in order.  Quite properly, tremendous effort has
been put into measuring neutrino masses.  Yet the precise
values of
neutrino masses are no more fundamental than light quark masses, and
they are much less important for the structure of the world.  In any case,
it is
unlikely we will reach a profound understanding of one without the other, since
in unified theories quarks and leptons come as a package.

Besides this general sort of motivation, there are several more specific
reasons to
be  especially interested in the values of  $m_u$ and $m_d$.   We would
like to be
absolutely sure that $m_u \neq 0$, since this empowers the P- and T-violating
$\theta$ parameter, which provides the prime motivation for Peccei-Quinn
symmetry and axions, with their profound physical and cosmological
implications.
The ratio $m_u/m_d$ partially determines the axion coupling, and will play
a vital
role in pinning down the underlying microscopic parameters if and when axions
are observed.   The proton-neutron mass difference, whose value makes all the
difference for nuclear stability and for stellar and cosmic nucleosynthesis
-- and
thereby to the structure and composition of matter --   of course depends quite
directly upon $m_u -m_d$.   Finally, there is the prospect of making precise
quantitative predictions for appropriate measurable isospin-violating
processes.
Especially promising in this regard are transitions between heavy quark-heavy
antiquark states by soft $\pi^0$  emission.   Conversely, these processes
allow a different, and
potentially cleaner, path to the determination of $m_u -m_d/m_u +m_d$ than
the traditional ways
involving electromagnetic mass differences or light hadron processes.

{\protect\boldmath$m_s$\qquad}
The strange quark mass plays a singular role in QCD.  For all the other quark
masses, it makes good sense to expand either in $m/\Lambda$ or $\Lambda/m$,
where $\Lambda$ is the primary QCD scale, vaguely in the neighborhood of
300 MeV.   For the strange quark neither expansion is clearly appropriate.  A
fundamental question  that I find quite interesting, and which is quite
important
for the interpretation of numerical work on light hadron spectroscopy, is
how this
spectroscopy depends on $m_s$.  Specifically, how does the ratio-sequence
$f_\pi:m_\rho:m_N:m_\Delta$ vary as $m_s$ is taken from infinity down to
zero?
The (un?)reasonable success of the valence-quark/bag model, and of the
quenched approximation, seem to suggest very mild dependence, but the
instanton liquid model seems to suggest more dramatic effects.

There are several major issues in QCD, especially regarding its phase
structure,
that hang on the value of $m_s$.  I'll mention them in due course.

\subsubsection{Weak Matrix Elements that\\ Add Value to Experiments}

The precise determination of $\epsilon^\prime/\epsilon$ in $K$ meson
decays, and
the cornucopia of results on $B$ meson decays and CP violation emerging from
the BABAR and BELLE collaborations, are beautiful and outstanding achievements
in experimental physics.   But to extract this work's full potential we'll
need to
produce theoretical calculations of hadronic matrix elements with comparable
accuracy.   Only after controlling this ``QCD background'' can we make
inferences
regarding the precise values of weak mixing angles and phases, and the possible
influence of physics beyond the standard model.

\begin{figure}[ht]
\centerline{\BoxedEPSF{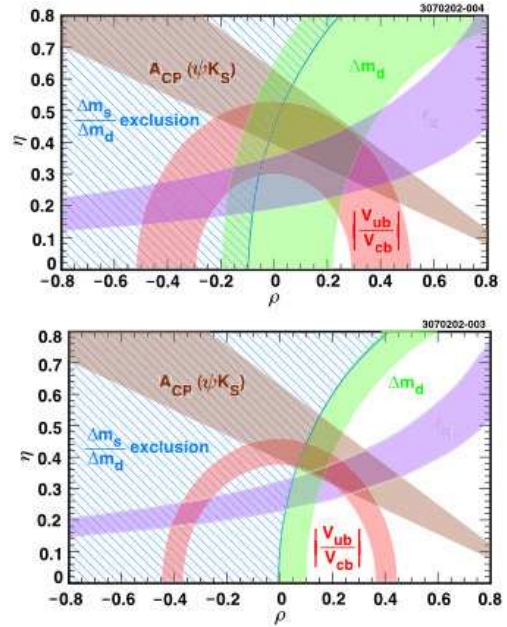 scaled 2300}}
\vspace*{-1.5pc}
\caption{Constraints on the standard model parameters $\rho$ and $\eta$ (one
sigma confidence level.  For the standard model to be correct, they must be
restricted to the region of overlap of the solidly colored bands.  The
upper figure
displays the constraints as they exist today.  The lower figure displays the
constraints as the would be if the precision of the lattice gauge theory
calculation
were reduced to 3\%. Figures provided by R. Patterson, Cornell University.}
\label{F2}
\vspace*{-\smallskipamount}
\end{figure}

Here there is a picture that is worth a thousand words.   Figures~\ref{F2}a
and~\ref{F2}b both depict allowed regions for a variety of experimental
measurements whose results depend on the weak mixing parameters $\rho,
\eta$, in the Wolfenstein  parametrization of the CKM matrix.    There is a
consistent determination, and no evidence for physics beyond the standard
model, in the region of overlap.   Figure~\ref{F2}a is the contemporary
situation;
Figure~\ref{F2}b assumes the same data, but with the theoretical precision
improved by a few teraflop-years of computation.   You can see the value
added.

\subsection{Fantasies}
\medskip
\subsubsection{Pinning Down \protect\boldmath$(g-2)_\mu$}

Another recent heroic, beautiful, and outstanding achievement in experimental
physics is the precision measurement of  the anomalous magnetic moment of the
muon, $(g-2)_\mu$, reported by the Brookhaven $g-2$ collaboration.   In
this case
too, the significance of this result would be greatly enhanced if we could do a
better job on the QCD ingredients, here low-energy vacuum polarization and
light-by-light scattering.    Our inability to do these calculations brings
home the
general point that the existing technology of lattice gauge theory is quite
fragile, in the sense that it can be used to calculate a few things very
well, but
most others essentially not at all.    So my fantasy here is of techniques that
degrade more gracefully.

Actually, in the specific case of $(g-2)_\mu$, numerical work might supply an
important input even while falling well short of the ``fantasy'' of
first-principles
calculation.     At present the major uncertainty in the standard model
prediction
arises from a discrepancy between the value of the hadronic vacuum polarization
as extracted extracted from $e^+e^-$ annihilation or from $\tau$
decay.   Even a fairly crude numerical determination of the vacuum
polarization at a single Euclidean
momentum might tell us which to believe.

\subsubsection{Quantitative Anthropics}

At the frontiers of where physics has come under control, we can start to
contemplate expansion into metaphysics.   Specifically, knowing the continuous
parameters (in mathematicalese, moduli) of our world-theory, we can ask -- and
hope to answer! -- a precise scientific  version of the great question:
Could things
have been different?  It seems to me that this question has taken on a new
urgency due to recent developments in physics, as I have discussed at length
elsewhere.    There are two important refinements of the question:

\begin{itemize}

\item{\it Anthropic Principle}:  Are the values of some or all of the moduli
determined by the requirement that there should be intelligent observers
capable of detecting them?   This could be true if the Universe were
inhomogeneous on ultra-large scales, with variable values of the moduli.   Then
the principle would emerge as a sort of converse to natural selection,
selecting out the environments to which intelligence might successfully
adapt.

\item{\it Misanthropic Principle}:  Are the values of some or all of the moduli
determined by historical accident?  This  could arise in the same
circumstances.  It
is the analogue of genetic drift.

\end{itemize}

If we could make a strong case that small variations in $m_u$ and $m_d$, to be
specific, would destroy the possibility of intelligent observers, then that
would be
{\it prima facie\/} evidence for the relevance of the anthropic principle,
since
these quantities appear to be very remotely conditioned by central
principles of
unified field theories, string theory, or anything else.

Fortunately, this subject has an empirical side.  We can search for
variation in the
physical ``constants" as a function of location or of time.   For
theorists, there is the
challenge to quantify what we learn about possible changes in the fundamental
constants from cosmic nucleosynthesis and from measurements of hyperfine
splittings at different times, for example.  One of the most powerful
constraints on
variation of constants comes from the Oklo natural reactor, which would
have left
a different residue had an accidental near-degeneracy in the levels of
Sm$^{149}$
+ n and Sm$^{150}$ not been present two billion years ago!  Clearly some of the
requisite calculations enter further into the realm of fantasy than others.
But it is a
valid and I think profound general observation that the freedom to do numerical
experiments with unrealistic values of moduli could help bring this particular
branch of metaphysics into the realm of hard science.

\section{Many-Particle Physics}

\subsection{Opportunities}
\medskip
\subsubsection{The \protect\boldmath$T\neq 0$ Equation of State}

A major achievement of lattice gauge theories in recent years has been to map
out the energy and pressure, as a function of temperature, for various
idealizations of QCD.   It is quite striking how the
number of effective degrees of freedom ascends to something near the free
quark-gluon value, at a remarkably low temperature by the standards of typical
hadron masses.   This phenomenon is one of the main inspirations for
experimental programs to study quark-gluon plasma at RHIC and eventually
ALICE.

A fully realistic simulation, with good implementation of chiral symmetry and
accurate values of the quark masses, is called for.   The value of the
strange quark
mass is especially important.  It is likely that for sufficiently heavy
strange quarks
there is a first-order phase transition, whereas below some threshold value
there
is only a crossover.   A very specific, concrete, and achievable goal is to
locate the
critical strange quark mass, and to verify (or not) existing indications that the
physical strange
quark mass is subcritical.  It is also important to quantify the
differential strange
quark contribution to pressure and density.   This would be an interesting
measure of how nearly free the quarks in the plasma are, and could provide a
baseline for interpreting enhanced strangeness multiplicities observed in heavy
ion collisions.

\subsubsection{Exploring Small \protect\boldmath$\mu/T$}

Finite chemical potential has long been {\it terra incognita\/} for numerical
simulation of QCD, because configuration by configuration the usual path
integral for the partition function contains a complex phase, and there are big
cancellations.

These cancellations are less severe for small $\mu/T$ and not too large
volumes, and are absent for imaginary $\mu$.
Exploration of these regimes has begun, but much more remains to be done,
especially in working towards realistic quark masses and reliable error
estimates.  There is a
potential connection to heavy ion experiments, where the effective chemical
potential varies with rapidity.

\subsection{Challenges}

\medskip

\subsubsection{Locate the True Critical Point}

There are good reasons to think that at zero temperature, as one varies the
chemical potential, there is a first-order quantum phase transition between
nuclear and quark matter.
There may in fact be several transitions of various kinds, including meson
condensation and alternative pairings in color superconductivity.   I
expect that most of these are
essentially low-temperature phenomena, leaving only a single first-order
transition above $\sim 50$ MeV.  In any case, for purposes of discussion
let me assume this.
Within this context one discovers a somewhat unconventional but I think
sharp and insightful perspective on the definition of ``quark-gluon
plasma''.  At sufficiently high (but not too high) chemical potential, as
one increases the temperature, or at sufficiently low  (but not too low)
temperature, as one increases the
chemical potential, there is a sharp phase transition.  We are justified in
calling this a hadron-to-quark transition, since the respective phases on
either side evolve without further
discontinuities into matter accurately described as nearly free hadrons or
nearly free quarks.

As the
temperature rises the distinction between hadronic and quark matter becomes
less  significant; the discontinuities associated with the first-order
transition grow smaller and
eventually vanish altogether.   The precise point in the
temperature-chemical potential plane where the distinction vanishes is
called the tricritical point.  It is associated with a
second-order phase transition, and critical fluctuations.   There is some
chance that the effects of such fluctuations could lead to observable
consequences in heavy ion collisions.

The challenge for lattice gauge theory is very clear and concrete: locate
this point!   It is a notable landmark in the QCD phase diagram, and
knowing its location would be very
helpful for organizing the experimental exploration of that diagram.

There is probably a close relationship between this physical tricritical
point and a more theoretical tricritical point implicit in our earlier
discussions.   That one arises in
considering behavior in the temperature-$m_s$ (strange quark mass) plane.
For very small $m_s$ one expects a first-order transition, which weakens as
$m_s$ increases, and
vanishes altogether at some critical $m_s^{\rm crit.}$.    A very
interesting possibility is that the theoretical tricritical point evolves
into the physical one, moving off into
nonzero values of $\mu$ above $m_s^{\rm crit.}$.  If so, it ought to be
possible to locate the physical tricritical point for values of $m_s$ just
above $m_s^{\rm crit.}$
by exploiting small $\mu/T$ techniques.

\subsubsection{Measure Critical Behavior}

The renormalization group analysis of QCD Lite -- that is, $SU(3)$ color
gauge theory with two massless
quarks -- suggests the existence of a second-order phase transition at
finite temperature, associated with
the restoration of chiral symmetry.   Universality arguments suggest that it
should exhibit the critical
exponents of a 4-component magnet, governed by the $O(4)$ linear $\sigma$
model.   There are many
precise predictions for the singular behavior of effective meson masses, specific
heat, magnitude of the
condensate, etc., based on this picture.   One can go on to predict
analytically the equation of state near
the critical point, also allowing for small but nonzero -- conceivably,
even realistic -- quark masses.

It would be a landmark achievement to check some of these predictions, say
specifically to measure a
critical exponent that can be clearly distinguished from mean field theory.    
This would provide welcomeconfirmation of some of our deepest prejudices about the nature of chiral
symmetry and its restoration; conversely, any deviation from the predictions would force us to rethink
some fundamentals.

A few years ago some doubts were raised about the applicability of
universality in this context.  They
have been convincingly addressed, but live on in modified form as valid
questions about the scope
of universality that deserve to be addressed quantitatively.  While now no
one doubts that there is universal
behavior near the critical point, it remains to quantify how wide the
critical region is, and how important
are the critical fluctuations against the background of conventional
thermodynamic behavior.   These are abstract but precise forms of 
the very basic question, What is the hadronic fluid like near the phase transition?
Is it dominated by color gauge
fields (glue degrees of freedom), so that the collective behavior of the
critical $\pi$ and $\sigma$ fields are a
minor side-show?  Or is it best pictured as a strongly interacting meson
gas?  In the former case, the critical
singularities will be blips on a smooth background; in the latter case,
they will be fractionally large or even
dominant.  The location of the critical point ($T_c$ well below glueball
masses, or even the scale set by the
QCD string tension) and that it is well below the $T_{c^\prime}$
for deconfinement in the pure
glue theory, suggest to me that the latter result is more likely, but I'm
not at all certain.

To do justice to these problems, one must both respect the chiral symmetry
and work in large spatial
volumes, so as to allow the appropriate modes to exist and have their
proper scope.  This will be very
costly to do directly, so use of a real-space (infrared!) renormalization
group or some other
novel technique might be necessary.

After this is done, it will still remain to address the related but
perhaps still more challenging
challenge of measuring the {\it tricritical\/} behavior.  They are expected
to be governed by mean field
theory, up to calculable logarithmic corrections.   In addition there are 
crossover exponents and a critical
equation of state awaiting measurement.

\subsection{Fantasies}
\medskip
\subsubsection{Reinvent Nuclear Physics}

The historical origin of strong-interaction physics, and its main
appearance in the natural world, is of course the physics of atomic nuclei.
Although with modern QCD we have achieved an extraordinarily beautiful and
``complete'' theory of the strong interaction, this fundamental progress
has not greatly
advanced our understanding of nuclei.  There is an obvious, good reason for
this.  The energy scales of interest in nuclear physics are of order a few
MeV, while the basic QCD
scale is a hundred times this.  So essentially nuclear phenomena are
governed by the residua of complicated cancellations among much larger
basic QCD forces.  It
is probably unrealistic, and unfruitful, to contemplate a brute force
assault from first principles, since small errors in large cancelling
quantities will dominate the
computed answers.

But it would be dereliction of duty, and a lost opportunity, to abandon the
field entirely.   A major goal of lattice gauge theory should be to compute
the parameters of effective
field theories, formulated in terms of pion and nucleon degrees of freedom,
from first principles.  (Then the effective theories could be turned over
to many-body
theorists.)  Specifically, we should verify that the theory generates such
key features as the hard core and the spin-orbit force, which are central
to qualitative aspects of nuclear
physics.   And we should get a convincing qualitative explanation of these
effects, backed up by exploration of QCD variants (see below).

It would also be amusing, and instructive, to explore the limits of
``nuclear'' physics as we know it.  Why are there nuclei at all?  In other
words, Why does the nuclear force
saturate?  Or in modern terms, Why do protons and neutrons in nuclei retain
their individual identity, rather than agglomerating into a single shared bag?  
Does this occur for two-color
QCD, or for supersymmetric QCD? -- maybe not, since in both these cases there are bosonic baryons.   
Does it occur in a world without light quarks?  In another direction:
How far is our world from a (literally) strange world, in which the ground
state for baryonic matter has nonzero strangeness?  How much lighter would
the strange quark have to
be?

\subsubsection{Reinvent Extreme Astrophysics}

There is a chance to break new ground, and to advance our understanding of
some of the most violent and crucial events in cosmic evolution, by getting
a better handle on the behavior of hadronic
matter in extreme conditions.   Contemporary astrophysical modeling of
supernova explosions and of the structure and evolution of compact stars is
largely based on poorly controlled
extrapolations of models abstracted from nuclear and resonance physics well
beyond their region of validity, into regimes where the particles are
strongly interacting or even overlapping.  In other
words, it is sophisticated guesswork.   We have good asymptotic theories
based on QCD for the highest temperatures and densities, but there is much
uncertainty about where and how asymptopia
is approached.   We may look forward to a wealth of relevant new data from
x-ray, neutrino, and gravitational wave detectors, in addition to dramatic
enhancements of the traditional optical
and radio windows.   Can we do justice to this information?  Specifically,
can we devise concrete signatures for quark matter and color
superconductivity in ``neutron star''  interiors, or
for phase transitions during or in the immediate aftermath of supernova
explosions or ``neutron star'' mergers?

\section{Conceptual and Algorithmic Issues}

\subsection{Opportunities}
\medskip
\subsubsection{See the Effects of Chiral Fermions}

Thanks to some very ingenious recent work we now know how to implement
chiral symmetry in a precise way in lattice QCD, suitable for numerical
integration.   As a result many
fundamental questions become accessible, including
\begin{itemize}
\item quantitative spectroscopy of the pseudoscalar mesons.  Since the
masses of the octet are most sensitive to the values of small light quark
masses, the meson masses can be
used in principle to determine the light quark masses, which then become
inputs to calculations of flavor $SU(3)$ breaking and isospin violation,
including ``electromagnetic''
mass differences;
\item questions of phase structure.  As has already appeared in our earlier
discussions, chiral symmetry restoration is a major feature
distinguishing different phases, and conditions the properties of
collective modes at the phase transitions,  it is important in these
contexts to implement it accurately;
\item study of axial baryon number violation.   Axial baryon number (more
precisely, axial $N_u +N_d$) is broken
intrinsically in the quantum theory by an anomaly, and to a small extent by
quark masses, and spontaneously by $\bar q q$ condensation.   Simulations
with large artifactual quark
masses mangle this quasi-symmetry, just as they mangle chiral $SU(2)\times
SU(2)$.    For all the associated questions of $\eta^\prime$ phenomenology,
squeezing out of
instantons and approximate $U(1)_A$ restoration at high temperature, and
axion cosmology, accurate implementation of {\it classical\/} $U(1)_A$ is
essential;
\item tests of the instanton liquid model.  This model predicts significant
dependence of certain hadronic quantities on light quark masses.
\end{itemize}

\subsubsection{Explore QCD Variants}

Besides enabling the noble bread-and-butter work of computing properties of
real-world QCD, lattice gauge theory offers us the opportunity to explore
variants, for example with
different quark mass spectra, different representations, or different gauge
groups.   I have already mentioned many possibilities for the creative use
of such flexibility, in
quantitative anthropics, in testing our understanding of the phase
structure, and in testing instanton liquid ideas.   This list is far from
exhaustive.  As a general remark, we can
test proposed qualitative explanations of phenomena in QCD by seeing
whether they predict the correct direction of change in the phenomena as
the underlying parameters
change.   For example, a question I find fascinating is simply: Why does
the naive quark model work so well?  In thinking about this question, it
would be very useful to identify
QCD-variants where the naive model begins to fail!   (Perhaps simply QCD
with many light quark flavors?)

Let me just mention one further potential application, to clear up a
long-standing debate whose origins are in the prehistory of QCD but that
remains contentious even now.  This
is the question of glueballs.  When a particle is observed experimentally,
it does not come labeled ``quark-antiquark'' or ``glueball'', nor is there
any strict  objective criterion to
distinguish them.  The whole distinction is tied up with the naive quark
model, which has no firm basis in quantum field theory, and ignores the
inevitability of mixing.
Nevertheless there is a simple objective question correlated with this
classification: If we vary quark masses, how does the mass of the particle
respond?

\subsection{Challenges}
\medskip
\subsubsection{Reduce Lattice Degrees of Freedom}

Full-scale QCD simulations tend to be unwieldy, inflexible, and opaque.
One would like to have procedures that fall somewhere between full-scale
simulation of the microscopic theory and rough
semiphenomenological modeling.

There are several suggestions regarding what are the important degrees of
freedom that could be used as the basis for a reduced description of QCD
that keeps contact with the microscopics,
including central vortices, abelian projections, instantons, and strings.
The challenge is to promote such proposals into reliable quantitative
tools, or at least to identify some limit in which they
become good, systematic approximations.

Alternatively, more directly numerically based approaches might rise to the
challenge.  The original Wilsonian program of the renormalization group, to
integrate up to a coarse lattice, and then
solve the coarse lattice theory by other means (presumably, some form of
strong coupling expansion) might be revisited in light of the vast increase
in knowledge and computer power over the last 25
years.

\subsubsection{Empower Reduced\\ Continuum Theories}

Extremely useful and important phenomenologies have been constructed using
reductions of QCD.   They are based in one way or another on integrating
out hard degrees of freedom, using
asymptotic freedom.   The primeval example is the analysis of deep
inelastic scattering using the operator product expansion, which reduces
what would otherwise be an entirely hopeless
problem to the evaluation of a discrete series of low-energy operator
matrix elements, or alternatively the structure functions of which they are
the moments.    Other applications, similar in spirit,
bring in Isgur-Wise functions and fragmentation functions.      Many of
these are poorly determined, yet they are important for the interpretation
of frontier experiments.  It would be a major
service, as well as an intellectual achievement, to compute more of them from
first principles.

\subsection{Fantasies}
\medskip
\subsubsection{Expand the  Frontiers of\\ Effective Computation}

Throughout most of this talk I have been discussing how numerical work can
address quantitatively demanding and/or sophisticated questions about a
particular strongly
interacting quantum field theory, QCD, or slight variants of it.    Of
course, at present this is the only precisely defined, strongly coupled
quantum field theory with proven
0relevance to the description of Nature.    But a lot of brainpower has been
expended on special limits of gauge theories, more or less distant cousins
of realistic QCD, for which
simplifications occur and some nonperturbative results can be obtained
analytically.  Notable among these are large $N$ limits and theories with
various degrees of
supersymmetry.   These and many other theories have been discussed in the
context of attempts at unification, as possible catalysts of electroweak,
unified, or super symmetry
breaking.     It would be very desirable to have even relatively crude
results about general classes of gauge theories.   Ironically, the limits
that provide analytical simplifications
seem to be especially difficult to simulate numerically using currently
known techniques.  So an attractive fantasy is to imagine numerical
techniques capable of coping with
\begin{itemize}
\item large $N$ gauge theories,
\item chiral gauge theories, and
\item supersymmetric gauge theories.
\end{itemize}
Recently there has been remarkable progress in formulating chiral fermion
theories (gauging these symmetries is {\it not\/} straightforward) and
supersymmetric lattice gauge
theories.  The constructions are very  technical and delicate, and not as
general as one might hope for.    Furthermore, potential Monte Carlo
simulations suffer from the fermion
sign problem.  So the fantasy here is to develop any usable algorithm.  For
large $N$ the standard is higher.  In principle, we know how to do the
calculations.  But one would like
to have algorithms that somehow simplify, rather than becoming
increasingly unwieldy, as $N$ increases!

Perhaps needless to say, even bigger game would come into sight if one
could develop usable nonperturbative methods for dealing with fermions in
general, finite density, or
real-time dynamics.

\subsubsection{Define the Limits of\\ Effective Computation}

I find it disturbing that it takes vast computer resources, and careful
limiting procedures, to simulate the mass and properties of a proton with
decent
accuracy.
Nature, of course, gets such results fast and effortlessly.  But how, if
not through some
kind of computation, or a process we can mimic by computation?  We are
accustomed to the idea that simulation of complex systems, or systems with
many elementary parts,
may be slow.   But here we are dealing with the simplest of objects in an
ideally simple physical theory.  Of course the underlying phenomenon is
that in quantum
field theory many interacting degrees of freedom are in play, even in the
simplest of physical circumstances.

Does this suggest that there are much more powerful forms of
computation that we might aspire to tap into?  Does it connect to the
emerging theory of quantum computers?  These musings suggest some concrete
challenges: Could a quantum computer calculate QCD processes efficiently?
Could it defeat the sign problem, that plagues all existing algorithms
with dynamical fermions?  Could it do real-time dynamics, which is beyond
the reach of existing, essentially Euclidean, methods?
Or, if all that fails,
does it suggest some limitation to the universality of computation?  There
is, after all, no guarantee that models abstracted from our (real or
imagined) implementations of
how to compute things  capture the process by which Nature decides how to
operate.  Maybe She just ``does it'', without computing anything.

A different sort of limitation to effective computation has become a major
theme of recent investigations in classical dynamics.   It is the ubiquity
of extreme sensitivity to initial
conditions, in problems ranging from the double pendulum to the long-term
behavior of the Solar System.   Are there hadronic properties that depend
sensitively on the
fundamental parameters we use to compute  them in QCD  (specifically, quark
masses), or on the small ``nonrenormalizable'' corrections due to other
interactions, or
discretization artifacts?   This is plausible, since in an effective
description we will have many channels with the same quantum numbers, or in
the language of dynamical systems
overlapping resonances, contributing to high orders in perturbation theory.
So even if the parameters of the effective theory depend smoothly on the
fundamental inputs,
physical quantities such as exclusive scattering amplitudes at intermediate
energies may not.   This would provide a rational explanation for the
terrible difficulties we've
had in devising reasonable algorithms for computing such quantities.  But
our understanding of questions like this is not even in its infancy.

\end{document}